\documentclass[aps,prb,twocolumn,groupedaddress,showpacs]{revtex4}
\usepackage{graphicx}
\newcommand{\be}{\begin{equation}}
\newcommand{\ee}{\end{equation}}
\newcommand{\bea}{\begin{eqnarray}}
\newcommand{\eea}{\end{eqnarray}}

\begin{document}

\title{ Quantum corrections of the Dzyaloshinskii-Moriya interaction on the spin-$\frac{1}{2}$ AF-Heisenberg chain in an uniform magnetic field }

\author{ S. Mahdavifar$^{1,2}$, M. R. Soltani$^3$ and A. A. Masoudi$^{4,5}$ }
\address{$^{1}$Department of Physics, Guilan University,
P.O.Box 41335-1914, Rasht, Iran\\
$^{2}$Institute for Advanced Studies in Basic Sciences, P.O.Box
45195-1159, Zanjan, Iran\\
$^3$Physics Research Center, Science Research Branch, Islamic Azad University,P.O.Box  19585-466,Tehran-Iran\\
$^4$Department of Physics, Islamic Azad university , Branch of North Tehran, P.O.Box  19585-936, Tehran, Iran\\
$^5$Department of Physics, Alzahra University, P.O.Box 19834, Tehran, Iran}
%%%%%%%%%%%%%%%%%%%%%%%%%%%%%%%%%%%%%%%%%%%%%%%
\begin{abstract}

  We have investigated the ground state phase diagram of the 1D AF spin-$\frac{1}{2}$ Heisenberg model with the staggered Dzyaloshinskii-Moriya (DM) interaction in an external uniform magnetic field $H$. We have used the exact
  diagonalization technique. In the absence of the uniform magnetic field ($H=0$), we have shown that the DM interaction induces a staggered chiral phase. The staggered chiral phase remains stable even in the presence of the uniform magnetic field. We have identified that the ground state phase diagram consists of four Luttinger liquid, staggered chiral, spin-flop, and ferromagnetic phases. 
  
  \end{abstract}
%%%%%%%%%%%%%%%%%%%%%%%%%%%%%%%%%%%%%%%%%%%%%%%%%%%
\pacs{ 75.10.Jm, 75.10.Pq}

\maketitle
%%%%%%%%%%%%%%%%%%%%%%%%%%%%%%%%%%%%%%%%%%%%

\section{Introduction}

   The effect of an external magnetic field on the quantum properties of the 1D antiferromagnetic (AF) spin-$\frac{1}{2}$ model has attracted much interest in recent years. Experimental and theoretical studies of this system have revealed a plethora of quantum fluctuation phenomena, not usually observed in higher dimensions. The Hamiltonian of this model in a uniform magnetic field ($H$) on a periodic chain of $N$ sites is given by  
%***********************************************************
\begin{equation}
\hat{H}= \sum_{j=1}^{N} [J \overrightarrow{S}_{j}.
\overrightarrow{S}_{j+1}-H S_{j}^{x}],\label{Hamiltonian1}
\end{equation}
%***********************************************************
where $J>0$ is the exchange coupling and $H$ is a uniform magnetic field. Theoretically, in the absence of the external magnetic field, $H=0$, the exact solution is given by the Bethe ansatz\cite{yang}. The spectrum is gapless and in the ground state, the system is in the Luttinger liquid phase, where the decay of correlations follow a power law. When a uniform magnetic field is applied the spectrum of the system remains gapless until the critical field $H_{c}=2 J$. Here a phase transition of the Pokrovsky-Talapov type\cite{pokrovsky} occurs and the ground state becomes a complete ordered ferromagnetic state\cite{griffiths}.
 
The progress in the experimental front is achived by introduction of high-field neutron scattering studies and synthesis of magnetic quasi-one dimensional systems. In many cases experimental data deviate significantly from the theoretical predictions based on the pure isotropic AF Heisenberg model in the uniform magnetic field\cite{dender, sirker, sakai, chaboussant, kageyama, cepas, jaime}. These deviations are due to anisotropies, most notably the Dzyaloshinskii-Moriya (DM) anisotropy\cite{dzyaloshinskii, moriya}. The Hamiltonian of this model is written as
%***********************************************************
\begin{equation}
\hat{H}= \sum_{j} [J \overrightarrow{S}_{j}.
\overrightarrow{S}_{j+1}+(-1)^{j} \overrightarrow{D}.(\overrightarrow{S}_{j}\times \overrightarrow{S}_{j+1})-H S_{j}^{x}],\label{Hamiltonian2}
\end{equation}
%***********************************************************
where $\overrightarrow{D}$ is the DM vector and the direction of this vector will be chosen along the $y$ axis, $\overrightarrow{D}=(0, D, 0)$. In actual systems, the direction of the $\overrightarrow{D}$ vector is fixed by the microscopic arrangement of atoms and orbitals. Since the DM interaction breaks the fundamental SU(2) symmetry of the isotropic Heisenberg interactions, it is at the origin of many deviations from pure Heisenberg behavior. Such anisotropy induces qualitatively different effects. In particular, in the 1D AF spin-$\frac{1}{2}$ model with the DM interaction a gap is opened in the energy spectrum and scales as $\Delta\sim (DH)^{\frac{2}{3}}$ in contrast with the pure Heisenberg case  (Eq.(\ref{Hamiltonian1})). Theoretically, using bosonization techniques Oshikawa and Affleck\cite{oshikawa} explained the observed scaling behavior of the energy gap. They have shown that in the presence of the staggered DM interaction along the chain, an applied uniform field $\overrightarrow{H}$ also generates an effective staggered magnetic field $\overrightarrow{h}\propto\overrightarrow{D}\times\overrightarrow{H}$. The staggered magnetic field for $H=0$ produces an AF ordered (Neel order) ground state and induces a gap in the spectrum of the model it is scaled as $h^{\frac{2}{3}}$. 
 
 For the higher-dimensional cases there is a theoretical expectation\cite{oshikawa, sato, fouet} that the field dependence of one of the gaps should be $\Delta \sim (D H)^{\frac{1}{2}}$. Fouet et al. also studied\cite{fouet} the gap-induced by the staggered magnetic field at the saturation uniform field $H_{c}=2 J$. Using field theoretical arguments and density matrix renormalization group (DMRG) method, they found that the gap scales as $\Delta(H_{c}) \sim D^{\frac{4}{5}}$. In a very recent work, this scaling behavior is clarified by using exact diagonalization Lanczos results\cite{mahdavifar07}. 
 Also, it was shown that in the case of the 2D frustrated dimer singlet spin systems, a magnetic field induces staggered magnetization\cite{kodama}. 

It should be noted that, most of the studies have excluded from consideration the quantum effects associated with the DM interaction\cite{oshikawa, sato, wang, alcaraz1,lou1}. In a recent work\cite{chernyshev}, the effect of an external magnetic field on the 2D AF Heisenberg model with DM interaction is studied. The dependence of the quantum corrections on the DM interaction is investigated. It is shown that the effect of the external field on the gap can be predicted by investigating the on-site magnetization of the model. On the other hand, the interplay of DM interactions and an external magnetic field in spin-$\frac{1}{2}$ dimers is studied\cite{miyahara}. It is shown that the staggered magnetization of an isolated dimer has a maximum close to one-half the polarization, with a large maximal value of $\sim 0.35$ in the limit of very small DM interaction. They have also investigated the effect of the inter-dimer coupling in the context of ladders with DMRG calculations. However the interply of the DM interaction on the ground state properties of the 1D AF spin-$\frac{1}{2}$ Heisenberg model is much less studied. Since the integrability of the model will be lost in the presence of the DM interaction, very intensive studies are needed.        
%***********************************************************
\begin{figure}[tbp]
\centerline{\includegraphics[width=8cm,angle=0]{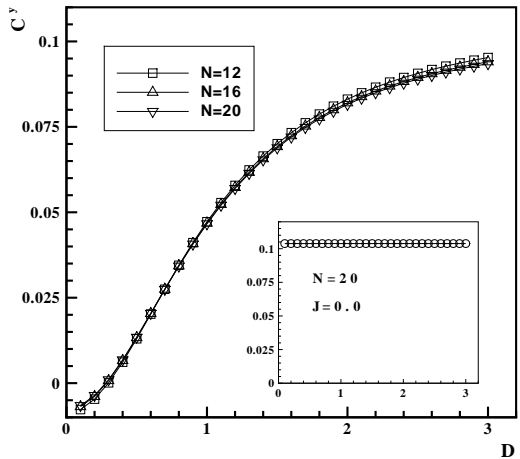}}
\caption{ The staggered chiral correlation function at $H=0$ plotted as a function of the DM vector $D$. The results reported for different chain lengths $N=12, 16, 20$ and $J=1$. In the inset, the value of the staggered chiral correlation function $C^{y}$ is shown versus $D$ for the pure DM interaction ($H=0$ and $J=0$) and chain length $N=20$. } \label{fig1}
\end{figure}
%*********************************************************** 

 In this paper, we present our numerical results obtained on the low-energy
 states of the 1D AF spin-$\frac{1}{2}$ Heisenberg model with the staggered DM interaction ($D$) in an external uniform magnetic field ($H$). We study the mutual effect of a uniform magnetic field and DM interaction on the ground state phase diagram of the model. In particular, we apply the modified Lanczos method to diagonalize numerically finite chains. Using the exact diagonalization results, we calculate the spin gap, the magnetization, the staggered magnetization, the staggered chirality and various spin-structure factors as a function of the uniform magnetic field ($H$) and DM interaction ($D$). Based on the exact diagonalization results, we obtain the ground state magnetic phase diagram of the model showing the Luttinger liquid, the staggered chiral, the spin-flop, and the ferromagnetic phases. We denote by "ferromagnetic phase" the phase with the magnetization parallel to the external magnetic field as only nonvanishing order parameter.
 
 The outline of the paper is as follows: In section II we present our numerical results of the exact diagonalization calculations on the ground state properties of the AF Heisenberg chain with the DM interaction. In section III we investigate the effect of a uniform magnetic field on the ground state properties of the model. Finally we conclude and summarize our results in section IV. 
 
%%%%%%%%%%%%%%%%%%%%%%%%%%%%%%%%%%%%%%%%%%%
%%%%%%%%%%%%Sec II%%%%%%%%%%%%%%%%%%%%%%%%%
\section{In the case of $H=0$}

In this section we explain the behavior of the model in the absence of the uniform magnetic field ($H=0$). Classically, the effect of a staggered DM interaction is interesting. 
The DM interaction makes it energetically favorable for the spins to stay in the plane perpendicular to the direction of $\overrightarrow{D}$ ($x$-$z$ plane). Without the $"J"$ term, the DM interaction makes the spins in different sublattices to be under a $\theta=\frac{\pi}{2}$ angle to each other. Thus at $H=0$, in the ground state of the model (2) spins lie in the $x$-$z$ plane (easy plane). 

On the other hand, Nersesyan et.al. predicted\cite{nersesyan} that in the anisotropyc (easy-plane) AF spin-$\frac{1}{2}$ chain with sufficiently strong frustrating next-nearest-neighbor coupling, a phase with a broken parity appears, which is 
characterized by the nonzero value of the chirality
%***********************************************************
\begin{eqnarray}
\chi_{j}^{\alpha}\equiv \langle (\overrightarrow{S}_{j}\times \overrightarrow{S}_{j+1})^{\alpha}\rangle,
\end{eqnarray}
%*********************************************************** 
where $\alpha$ denotes ($x, y, z$) and the notation $\langle...\rangle$ represent the expectation value at the lowest energy state. However, two different types of the chiral ordered phases, gapped and gapless were found\cite{kaburagi, hikihara}. In order to explore the nature of the spectrum and the ground state phase diagram of the model, we have used the modified Lanczos method\cite{grosso, langari1} to diagonalize numerically finite ($N=12, 14, ..., 24$) chains. The energies of the few lowest eigenstates were obtained for chains with periodic boundary conditions. 

We have calculated numerically the staggered chiral order parameter $\frac{1}{N}\sum_{j}(-1)^{j}\chi_{j}^{y}$ and the staggered chiral correlation function defined as 
 %***********************************************************
\begin{eqnarray}
C^{y}=\frac{1}{N} \sum_{n=1}^{N} (-1)^{n}\langle \chi_{j}^{y} \chi_{j+n}^{y}\rangle.
\end{eqnarray}
%*********************************************************** 
Lanczos results lead to $\chi=0$ for any value of the DM vector $D$, because in a finite system no symmetry breaking happens. In Fig.1 we have plotted the staggered chiral correlation function along the $"y"$ axis, $C^{y}$, as a function of DM vector $D$ for different chain lengths $N=12, 16, 20$. As can clearly be seen, the staggered chiral correlation function $C^{y}$, increases with increasing $D$, which shows that the DM interaction suppresses the quantum fluctuations in the $x$-$z$ plane and induces a staggered chiral phase in the ground state phase diagram of the model. Introducing a DM interaction, the SU(2) rotational symmetry breaks and a quantum phase transition happens in the ground state phase diagram of the model.
It is important to note that due to the profound effect of quantum fluctuations the chirality does not saturate. 
We have also checked the excitation energies of the three lowest levels as a function of $D$. We have considered the excitation gap in the system as the difference between the first excited state and the ground state. We have found that a gap opens in the presence of the DM vector $D$. In the inset of Fig.1, the staggered chiral correlation function $C^{y}$ has been plotted as a function of $D$ for the pure DM interaction ($H=0$ and $J=0$) and chain length $N=20$. It shows that in the absence of Heisenberg interaction ($J=0$) and Zeeman term ($H=0$), the staggered chiral order is governed by $D>0$.

Thus, in the absence of a DM interaction, the ground state of the system is in the gapless Luttinger liquid phase with a power-low decay of correlations. Adding a DM interaction to the isotropic Heisenberg model develops a gap. The ground state then has the long-range staggered chiral order in the $y$ direction. 

%***********************************************************
\begin{figure}[tbp]
\centerline{\includegraphics[width=8cm,angle=0]{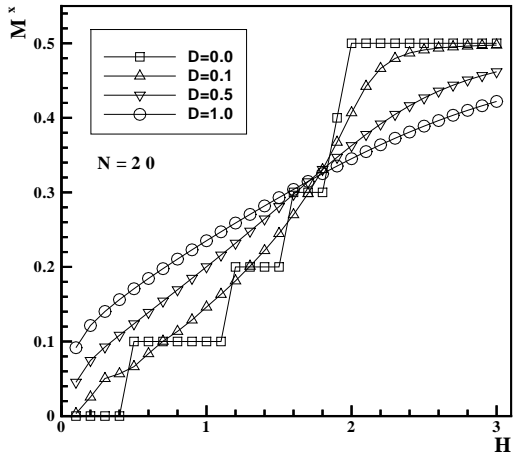}}
\caption{ The uniform magnetization $M^{x}$ as a function of applied magnetic field $H$ for $N=20$ chain for different values of the anisotropy DM vector $D=0.0, 0.1, 0.5, 1.0$. The value of $J=1$ is considered.} \label{fig2}
\end{figure}
%***********************************************************

 %%%%%%%%%%%%%%%%%%%%%%%%%%%%%%%%%%%%%%%%%%%%
 
 %%%%%%%%%%%  Sec III  %%%%%%%%%%%%%%%%%%%%%%%%%%%%%
 
\section{In the case of $H\neq 0$}

In this section we study the effect of a uniform magnetic field on the ground state phase diagram of the 1D AF spin-$\frac{1}{2}$ Heisenberg model with DM interaction. As we mentioned, in the absence of the uniform magnetic field ($H=0$) and DM interaction ($D=0$), the spectrum is gapless. The ground state is in the Luttinger liquid phase. By applying a uniform magnetic field $H$, the SU(2) symmetry of the pure Heisenberg model reduces to a U(1) symmetry corresponding to a rotation around the magnetic field direction ($x$ axis). The spectrum remains gapless until a critical saturation field $H_{c}=2 J$. As soon as a staggered DM interaction with a $\overrightarrow{D}$ vectore not parallel to the uniform magnetic field ($H$) is introduced, the rotational symmetry in spin space is completely lost. The only symmetry that remains is the mirror symmetry with respect to the $x$-$y$ plane (the plane contaning the uniform magnetic field and the $\overrightarrow{D}$ vector). As a consequence, the staggered magnetization per site must lie in the $z$ direction (the direction perpendicular to the plane defined by the uniform magnetic field and the $\overrightarrow{D}$ vector), while the uniform magnetization per site is in the field direction.
If the  $\overrightarrow{D}$ vector is parallel to the uniform magnetic field, the U(1) rotational symmetry is still present. Thus, the staggered magnetization is identically zero. In following we show that the quantum phase transitions can be easily observed from the numerical calculations of the small systems.

 The symmetry breaking considerations suggest that an insight into the nature of different phases can be obtained by studying the magnetization 
 %***********************************************************
\begin{eqnarray}
M^{\alpha}=\frac{1}{N}\sum_{j}\langle S_{j}^{\alpha}\rangle,
\end{eqnarray}
%*********************************************************** 
and the staggered magnetization  
%***********************************************************
\begin{eqnarray}
M_{st}^{\alpha}=\frac{1}{N}\sum_{j}(-1)^{j}\langle S_{j}^{\alpha}\rangle,
\end{eqnarray}
%*********************************************************** 
and the spin correlation functions. The static spin structure factor at momentum $q$ is defined as 
%***********************************************************
\begin{eqnarray}
S^{\alpha \alpha}(q)=\sum_{n}e^{i q n}\langle S_{j}^{\alpha}S_{j+n}^{\alpha}\rangle.
\end{eqnarray}
%*********************************************************** 
%***********************************************************
\begin{figure}[tbp]
\centerline{\includegraphics[width=8cm,angle=0]{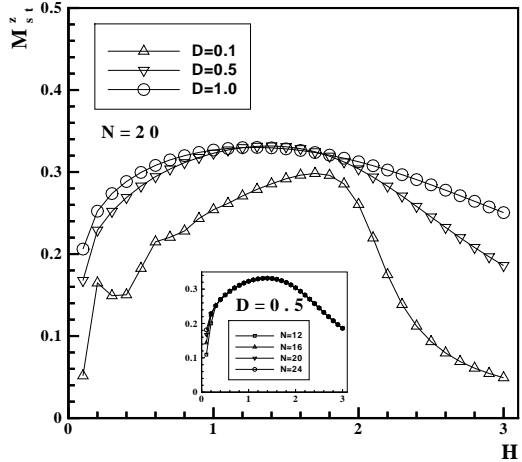}}
\caption{ The staggered magnetization along $z$ axis, $M_{st}^{z}$ as a function of applied magnetic field $H$ for $N=20$ chain for different values of the anisotropy DM vector $D=0.1, 0.5, 1.0$. The value of $J=1$ is considered.} \label{fig3}
\end{figure}
%***********************************************************
It is known that the spin structure factors give us deeper insight into the characteristics of the ground state. In particular we study the $H$-dependence and $D$-dependence of different spin stucture factors. To determine the properties of this model in different sectors of the ground state phase diagram we have implemented the Lanczos algorithm of the finite chains $N=12, 14, ..., 24$ to calculate the lowest energy state. We have computed the ground state for different values of the DM vector $D=0, 0.1, 0.5, 1.0$. In Fig.2 we have plotted the magnetization along the applied uniform field, $M^{x}$ versus $H$ for chain length $N=20$ and different values of the anisotropy $D$. It can be seen, that for the $D=0$, due to the effect of the quantum fluctuations in finite sizes the magnetization remains zero for smal values of the uniform magnetic field $H$. For $H>H_{c}=2 J$ the magnetization saturate. This is in agreement with results obtained within theoretical approaches. Due to the quantum fluctuations, in the presence of $D$ there is no sharp transition to the saturation value of the magnetization. When $D\neq0$, the magnetization develops as soon as the magnetic field is swiched on, only reaching saturation asymptotically in the limit of infinite field. We mensioned that at the $H=0$ the staggered DM interaction causes the spins stay in the $x$-$z$ plane. In this case, the effect of the uniform magnetic field decreases the degeneracy of the ground state energy. 
%***********************************************************
\begin{figure}[tbp]
\centerline{\includegraphics[width=8cm,angle=0]{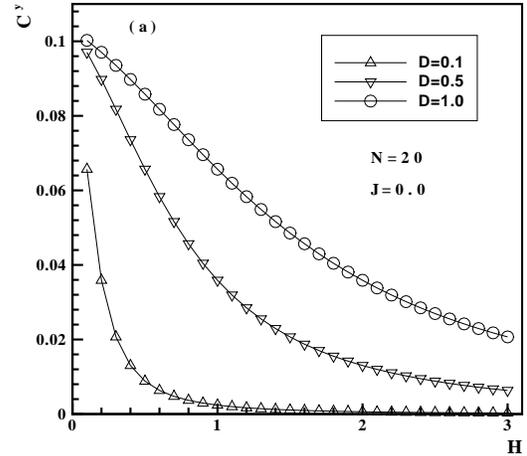}}
\centerline{\includegraphics[width=8cm,angle=0]{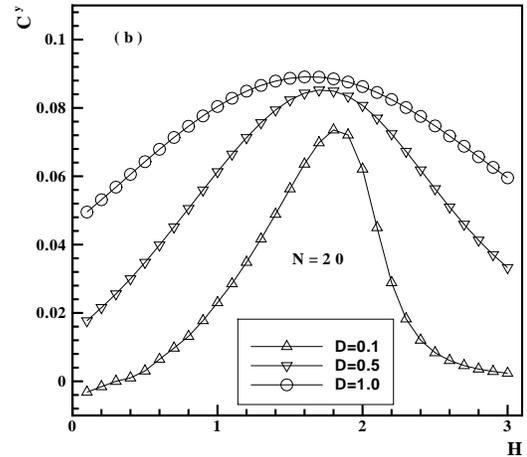}}
%\centerline{\includegraphics[width=8cm,angle=0]{fig4a.eps}}
\caption{ The  value of the staggered chiral correlation function along $y$ axis, $C_{y}$ at the fixed different DM vectors $D=0.1, 0.5, 1.0$ versus the uniform magnetic field $H$ for (a) $J=0$ and (b) $J\ne 0 (J=1)$. The results reported for a chain length $N=20$.} \label{fig4}
\end{figure}
%***********************************************************

In Fig.3 we have also plotted the staggered magnetization along the $z$ axis $M_{st}^{z}$, as a function of the uniform magnetic field $H$. The results reported for a chain length $N=20$ and different DM vectors $D=0.1, 0.5, 1.0$. It shows in complete agreement with the theoretical results of the effective Hamiltonian\cite{oshikawa} and symmetry breaking considerations, by applying a uniform field $H$, a profound Neel order in the $z$ direction induces. Which shows that there is long range spin-flop order along $z$ axis. The oscilations of $M_{st}^{z}$ at finite $N$ for smal values of DM vector $D$, are the result of level crossing between ground state and excited states of the model. There is also a maximal value for the staggered magnetization per site, around $\sim 0.3$. This value is of the order of the maximal value of the isolated dimer, and it depends relatively weakly on $D$. The inset of Fig.3 shows the staggered magnetization along the $"z"$ axis versus the uniform magnetic field $H$ at DM vector $D=0.5$ and different chain lengths $N=12, 16, 20, 24$. It can be seen that the maximal value of the staggered magnetization is independent of the system size. On the other hand, we have also investigated, $S^{zz}(q=\pi)$ as a function of the uniform field $H$ for different DM vectors. We have found that, there is a trend toward staggered magnetization along $"z"$ axis for $H>0$. Which confirms that the ground state of the model has the Neel long range order along the $"z"$ axis, which is known as the spin-flop phase.  
%***********************************************************
\begin{figure}[tbp]
\centerline{\includegraphics[width=8cm,angle=0]{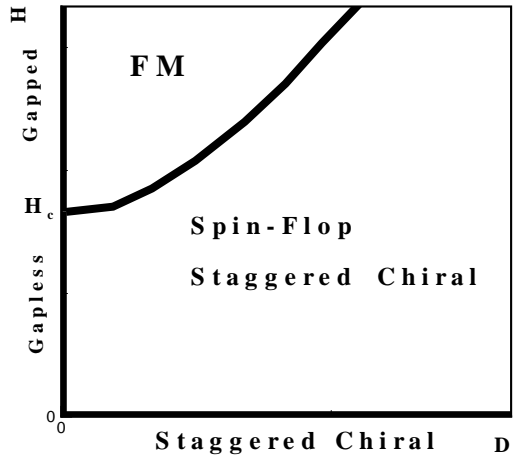}}
\caption{ Schematic picture of the ground state phase diagram of the 1D AF spin-$\frac{1}{2}$ Heisenberg model with DM interaction in an external uniform magnetic field ($H$).} \label{fig5}
\end{figure}
%***********************************************************

An additional insight into the nature of different phases can be investigated by studying the staggered chiral correlation function $C^{y}$. In this case we study the magnetic field dependence of the function $C^{y}$ for several values of the DM vector. 
In Fig.4a we have shown the field dependence of the staggered chiral correlation function $C^{y}$ for the pure DM interaction ($J=0$) and chain length $N=20$. As we mensioned before, in the absence of the Heisenberg interaction and Zeeman term, the ground state of the model has the long-range staggered chiral order along the $y$ direction. It can be clearly seen, that the staggered chiral phase remains stable even in the presence of the uniform magnetic field less than some critical field. Here a quantum phase transition occurs and the ground state becomes  a completely ordered ferromagnetic state. Our numerical results also show that the value of the critical field depends on the DM interaction. 

To obtain a complete picture of the ground state phase diagram of the model, we have also calculated the function $C^{y}$ for the Heisenberg model with the DM interaction. In Fig.4b the staggered chiral correlation function is plotted as a function of the uniform magnetic field $H$ for different values of the DM vector $D=0.1, 0.5, 1.0$. The results reported for a chain length $N=20$. As is seen from this figure, $C^{y}$ first increases versus the uniform magnetic field $H$ and after passing a maximum decreases. Since the uniform magnetic field suppress the quantum fluctuations in the $x$ direction, thus the staggered chirality increases increases up to the saturation. However, with more increasing the magnetic field $H$, the staggered chirality decreases from the saturation value in well agreement with Fig.4b. Also, due to the quantum fluctuations, the staggered chirality dose not reach to zero for large DM vectors.

Based on the symmetry analyses and numerical calculations, we expect that the ground state phase diagram of the model (2) has been a form shown on Fig.5. Here the ground state phase diagram of a 1D AF spin-$\frac{1}{2}$ Heisenberg model with the staggered DM interaction in an external uniform magnetic field, is presented on the $H$-$D$ plane. The ground state phase diagram contains, besides the gapless Luttinger liquid and gapped ferromagnetic (FM) phases, the gapped staggered chiral and spin-flop phases. The gapped staggered chiral and spin-flop phases are realized only in the case of DM interaction ($D>0$). Each phase is characterized by its own type of the long-range order: the ferromagnetic order along the magnetic field axis in the FM phase; the Neel order along the $"z"$ axis in the spin-flop phase; and the staggered chiral order along the $"y"$ axis in the staggered chiral phase.

In principle, the DM interaction breaks the fundamental SU(2) symmetry of the pure isotropic Heisenberg interactions and also U(1) symmetry of the Heisenberg model in the presence of a uniform magnetic field. The only symmetry that remains is the mirror symmetry with respect to the plane containing the uniform magnetic field and the $\overrightarrow{D}$ vector ($x$-$y$ plane). 

%%%%%%%%%%%%%%%%%%%%%%%%%%%%%%%%%%%%%%%%%%%%
 
 %%%%%%%%%%%  Sec IV  %%%%%%%%%%%%%%%%%%%%%%%%%%%%%
 
\section{conclusions}

In this paper, we have investigated the ground state phase diagram of the 1D AF spin-$\frac{1}{2}$ Heisenberg model with the staggered Dzyaloshinkii-Moriya interaction in an external uniform magnetic field $H$. We have implemented the modified Lanczos method to diagonalize numerically finite chains. Using the exact diagonalization results, we have calculated the spin gap, the magnetization, the staggered magnetization, the staggered chirality and various spin-structure factors as a function of the uniform magnetic field and DM interaction. In the absence of the uniform magnetic field $H=0$, we have shown that the DM interaction induces the staggered chiral phase. We have also found that a gap opens in the spectrum of the model. Thus, we have concluded that the gapped staggered chiral ordered phase appears in the ground state phase diagram of the 1D AF spin-$\frac{1}{2}$ Heisenberg model for $D>0$. We have identified that the application of a uniform magnetic field induces a spin-flop phase in the ground state phase diagram of the model.Also, the staggered chiral phase remains stable even in the presence of the uniform magnetic field.  Finally, we have shown that the ground state phase diagram consists of four Luttinger liquid, ferromagnetic (FM), staggered chiral and spin-flop phases.  
%%%%%%%%%%%%%%%%%%%%%%%%%%%%%%%%%%%%%%%%%%%%%%%%%%%%%%%%%%%%%%%%%%%%%%
\section{Acknowledgments}
M. R. H. Khajehpour, J. Abouie,  R. Jafari, M. Maleki and F. Mohammad-Rafiee are thanked for helpful discussions. 

\medskip

%-----------------------------------------------------------------------------

\end{document}